# CONTRASTING THE EFFICIENCY OF STOCK PRICE PREDICTION MODELS USING VARIOUS TYPES OF LSTM MODELS AIDED WITH SENTIMENT ANALYSIS

**Varun Sangwan (RA1911003010036 – vs5019@srmist.edu.in)**

**Vishesh Kumar (RA1911003010212 – vk7879@srmist.edu.in)**

**Dr. Bibin Christopher(bibinchv@srmist.edu.in)**

**Department of Computer Science & Engineering**

**SRM Institute OF Science And Technology (Chennai)**

---***---

## Abstract –

Stock market forecasts are always attracting the attention of multiple analysts and researchers. Many Popular theories observe that the stock market is, in its essence, a random game of chance and that it is a mindless game to try to predict a thin item. Predicting a stock's price is a challenging problem due to the number of external variables present. The market behaves in the short term like a ballot or a voting machine, but in the long run, it acts like a weighing machine and may thus forecast market movements over a more extended time period. Stock Price prediction's integration with modern technology – especially Machine Learning Algorithms (Quant models as often referred to in the financial sector) is recently becoming a growing idea for research. Research has shown that Machine Learning Models, particularly with the use of Recursive Neural Networks (RNN) and Long-Term Short Memory (LSTM), when applied to historical data of shares, can be utilized to predict the short-term price of the share. Our research aims to find the best model that uses companies' projections and sector performances and how the given company fares accordingly to correctly predict equity share prices for both short- and long-term goals.

## 1.INTRODUCTION

Forecasting and analysing the stock market is attempting to predict the possible future value of an organisation'sorganisation's stock or other exchange-traded instruments of finance. The stock market is a critical aspect of the country'scountry's economy; it also plays a vital role in expanding its industry and trade, impacting its economy. Investors and the industry are interested in the stock market and want to know if particular stocks will vary over time. The stock exchange is the primary funding source for any firm seeking to grow its operations. It relies on the supply and demand idea. If there is a strong demand for the organisation's shares, the stock price rises; if there is little demand for the stock, the stock price falls.

The National Stock Exchange of India (abbreviated NSE) is India's principal stock

exchange, based in Mumbai. The NSE was established in 1992 as the country's first language-free digital exchange. The NSE is the first exchange in the country to provide a contemporary, fully autonomous screen-based online trading system, making it simple for investors across the country to trade. Similarly, various other stock exchanges like NASDAQ and DOW JONES in the United States of America (abbreviated as USA), NIKKEI in Japan, KOSPI in South Korea, FTSE in the United Kingdom(abbreviated as UK), DAX in Germany etc act as markets for securities to be bought and sold.

The main motive to correctly predict stock values and prices in the short and long term is to maximise your potential earnings rather than relying on tips. A significant amount of research has gone into developing Machine Learning models that can correctly predict stock prices and have been used by hedge funds and investment banks for quite some time now. However, these models are mainly used to predict short-term prices so that they can be utilised in intraday trading, and most long-term models generally focus on indices and option chains.

The efficiency of various prediction models can be debated as many can not predict long-term fluctuations and compare the current stock value as compared to its current trading price which takes into account the sector performance(For Eg: The existing share of a stock; assuming Reliance declines in the Oil Sector as compared to its projections signalling a down quarter, but its price value has not fluctuated much, leading to the assumption that a correction is in order which will result in large volumes of shares being sold and the stock price taking a hit).

Fundamental Analysis refers to the concept of using underlying financial records published by the company, taking other competitor data, and contrasting them to predict short- and long-term prices correctly. It requires using historical shares datasets with critical information like closing prices, volumes traded, uptrend, downtrend etc.

Because of the engagement of a wide number of sectors and enterprises, it has incredibly massive databases from which it is impossible to extract information manually and analyse working patterns. This project's application not only predicts the future movement of a stock in the market, but it also automates the retrieval of data, trend evaluation, predictive modelling, and insight production of a stock with the touch of a button.

Using sentiment analysis and NLP, a comparative study was done on the efficiency of various models to predict short and long-term share prices. Apart from vanilla LSTM and LSTM aided with sentiment analysis, other models were also implemented, like a Bidirectional LSTM, which is a sequence prediction model that consists of two LSTMS – one that runs in the forward direction and one that runs in the backward direction which aims to increase efficiency and reduce the margin for error.

The other models that have also been tested to find the most efficient model are Seq2Seq LSTM(an encoder-decoder model created using RNN) and the LSTM two-path approach. The use of Sentiment Analysis has also improved all these models. Sentiment analysis aims to remedy specific scenarios that cannot be predicted by numbers alone and are more dependent on real-world factors. For Eg: The Covid 19 Pandemic caused the sale of an unprecedented amount of shares dragging the markets down when everything was under lockdown leading to a collapse in a previously predicted up-trending market.

Stock market analysis and forecasts will reveal market trends and predict when to buy stocks. Successfully predicting the future price of a stock can generate substantial profits. Our project implements extensive training of 12-month historical market data for various company stocks like TESLA, Twitter, AMD, Facebook etc. , to represent various conditions and confirm that time series models have significant predictive power over the Statistical side for high probability trading and high return for competitive business investment.

## 2. LITERATURE SURVEY

[1] Research on Legitimate Neural Network Based Stock Price Prediction Method, IEEE 2019, authored by Sayavong Lounnapha et al. This research proposes to develop a stock price forecasting system based on sophisticated neural networks with exceptional self-learning capabilities. The dataset is taught and tested in terms of the CNN and Thai stock market price forecasts. Prediction accuracy is high and may be encouraged in the financial industry.

[2] Enhancing returns by predicting stock prices using Deep Neural Networks, IEEE 2019, authored by Soheila Abrishami et al. Financial Forecasting and Prediction is an enormous task that attracts the interest of several academics and is critical to investors. This research paper introduces a system using deep learning that predicts the value of a stock based on a sequence of data about a piece of a stock traded on the NASDAQ stock exchange. The model is trained using the minor data for a specific stock and correctly guesses its ultimate value in numerous phases. It incorporates an autoencoder for noise removal and employs time series data architecture to deliver improved features alongside the original features. These additional characteristics are also supplied into the stacked LSTM autoencoder, which estimates the ultimate stock value in many steps.

[3] LSTM Method for Bitcoin Price Prediction: A Case Study Stock Market Yahoo Finance, IEEE 2019 authored by Ferdiansyah et al. Due to the volatility of Bitcoin in the stock market, automated solutions are necessary. This research uses LSTM to provide bitcoin stock market prediction mode forecasts. Before validating the findings, the research attempts to quantify them using RMSE (square root squared error). The research finds that the RMSE will always be greater than or equal to the MAE. The RMSE measure assesses the model's ability to calculate continuous values.

[4] Stock Price Prediction Using Machine Learning Techniques, IEEE 2019, authored by Jeevan B et al. This research study focuses on stock price prediction on the National Stock Exchange utilising RNN (Regenerating Neural Network) and LSTM (Long Term Short Term Memory) employing several parameters. Current market values and anonymous incidents are examples of such causes. This article also discusses recommendation systems and models based on RNN and LSTM algorithms that are used to choose firms.

[5] Stock Market Prediction Using Machine Learning Techniques, IEEE 2020, authored by Naadun Sirimevan et al. This study uses behavioural reactions to web news to close the gap and make predictions far more accurate. A day, a week, and two weeks later, accurate forecasts were made.

[6] Stock Market Forecasting by Machine Learning, IEEE 2018, authored by Ishita Parmar, Ridam Arora et al. The application of regression and LSTM-based machine learning techniques for forecasting stock prices are investigated in this work. The elements measured are open, close, low, high, and volume. Using machine learning techniques, this research paper attempts to predict a company's future stock price with more accuracy and predictability. The LSTM algorithm produced a beneficial outcome with more accuracy in forecasting stock values.

[7] Stock Price Prediction Using Machine Learning, IEEE 2018, authored by Jeevan B, Naresh E et al. This research is primarily based on the action course prediction technique, which predicts the value of the action utilising long-term, short-term memory (LSTM) and recurrent neural network (RNN). Several elements, such as market price currents, price-to-earnings ratio, fundamental value, and other anonymous statistics, are used on NSE data. The model's performance was evaluated by comparing the real and predicted data using an RNN graph. Machine learning is used to forecast stock prices because the algorithm can predict prices extremely near the accurate price by capturing comprehensive features and employing various methodologies.

[8] Predictive Model Development for Stock Analysis, IEEE 2017, authored by R. Yamini Nivetha et al. The primary purpose of this research is to compare three algorithms: Multilinear Regression (MLR), Support Vector Machine (SVM), and Artificial Neural Network (ANN). The monthly and daily predictions will be used to anticipate the market price for the next day. Stock prices are predicted using sentiment analysis and the best prediction system. The multilinear regression technique is the least developed approach for calculating the volume and stock price association. The study's findings suggest that deep learning algorithms are more sophisticated than MLR and SVM algorithms.

[9] Stock Price Prediction Based on Information Entropy and Artificial Neural Networks, IEEE 2019 - Zang Yeze, Wang Yiying et al.

## 3. METHODOLOGY

The model was implemented on Google Colab and using python libraries like pandas, matplotlib, NumPy and yahoo finance. The objective was to predict stock prices using numerical, fundamental and sentiment analysis of companies. The data was first imported from the yahoo finance library of share prices which records all the changes in the stock prices on an interval of 5 minutes as provided by the share's stock exchange – NASDAQ, DOW JONES, NIFTY etc. In this particular testing model, we have used the Microsoft share and we have created the data frame based on the closing prices of the share daily and plotted them using matplotlib. A minmax scaler was then applied onto it for the purpose of rescaling all the values in the scale [0,1] and the reshaped model was then trained using Long Short-Term Memory and the model was then successfully compiled.

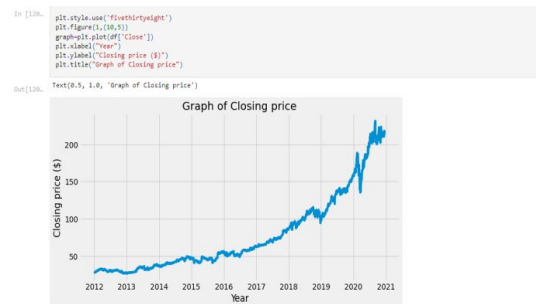

Fig 1: Closing price graph of Microsoft plotted

Two measurements are used to determine the genuine worth of the stock - The P/E ratio (Price/Earnings Ratio) The price/earnings ratio, also known as the price to earnings ratio or the P/E ratio, is a financial statistic that compares the price of a company's stock to its profits per share. Simply said, it depicts the relationship between stock price and earnings. We may use this ratio to determine how lucrative it is to purchase stock in a given firm. We may also use the P/E ratio to identify whether stocks are over or undervalued. For example, if two businesses in the same industry have entirely different P/E ratio values, it may indicate that the appraisal of one of them is not credible. The P/E ratio may be determined using the simple EPS (earnings per share) by dividing the current stock price by the earnings earned per share. The alternative metric is the price-to-sales ratio, abbreviated as the ratio. Because sales are sometimes referred to as a form of earnings, the P/S ratio is also known as price-to-earnings or price-to-earnings ratios.

It is mostly used to determine how much a stock is now worth in the market. This ratio is sometimes used in conjunction with the well-known P/E ratio to determine how appealing a firm is in comparison to its peers. The lower the price-to-sales ratio, the less cheap the firm appears to be, and the more "buy" the stock qualifies. For example, if firm Y has a price-to-sales ratio of 1.5 times that of company X, it might be argued that company Y is cheap. As a result, it makes sense to purchase business Y and sell firm X. While the P/S ratio is an excellent investing statistic, it is best to compare it to similar firms. For example, it makes no sense to compare the P/E ratio of a petrochemical company, like Shell, with a technology company, like Apple, because the two operate radically differently.

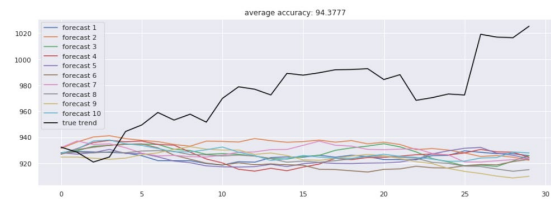

Fig2. Forecast values for vanilla-LSTM model

Using the following measurements, the data was scaled with a MinMax Scaler and prepared for testing before applying the LSTM principles. The Long-Term Short-Term Memory Network is a more sophisticated version of the Recurrent Neural Network, a sequential network that permits storing information. It can deal with the vanishing gradient problem that RNNs confront. A cyclic neural network, commonly known as an RNN, is a type of permanent memory. Units refer to the number of LSTM cells in the layer for the LSTM layer. The model will have a high dimensionality of 50 neurons, adequate to capture both upward and negative trends. Because we need to add another LSTM layer after the present one, return_sequences is set to True. The total amount of time stamps and indications is represented by input_shape. During each training cycle, 20% of the 50 neurons will be disregarded at random. The second, third, and fourth LSTM layers were added in the same manner as before. The data was then returned to its original size, and the model's predictions were compared to the actual closing price.

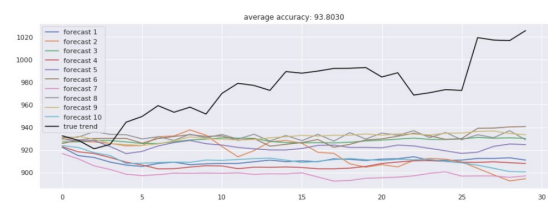

Fig3. Forecast values for Bidirectional-LSTM

Apart from normal LSTM, sentiment analysis other models were also implemented like a Bidirectional LSTM which is a sequence prediction model that consists of two LSTMS – one that runs in the forwad direction and one that runs in the backward direction which aims to increase efficiency and reduce the margin for error.

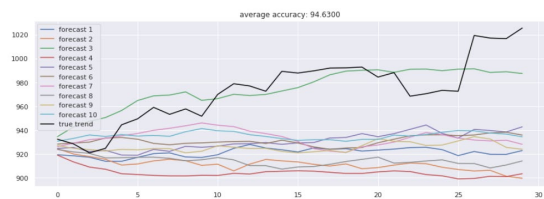

Fig4. Forecast values for LSTM 2-path model

The other models that have also been tested to find the most efficient model are Seq2Seq LSTM which is an encoder-decoder model made using RNN. It is useful in determining the trend of stock implemented in sentiment analysis on a database of respected papers and journals. Sentiment analysis is aimed to remedy certain scenarios that can't be predicted by number alone and are more dependent on real world factor. For Eg: The Covid 19 Pandemic caused the sale of an unprecedented amount of shares dragging the markets down when everything was shut down which led to a collapse in a predicted up-trending market.

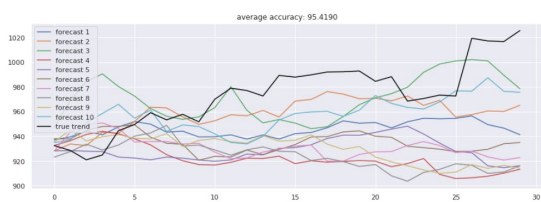

Fig5: Forecast values for LSTM seq-seq model

The model was also given input different stock datasets to generate buy-sell advisories on them based on the returns to volatility ratio.

## 4. RESULTS AND CONCLUSION

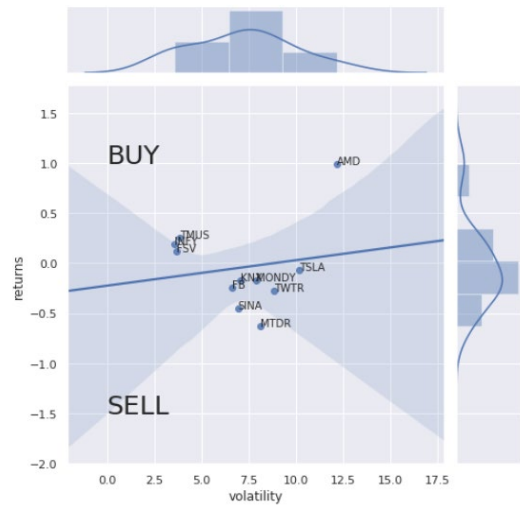

Fig5 : Stock advisory predictions based on a set of stocks given as inputs generated by the LSTM 2-path model

Our research pointed to the conclusion that the LSTM 2-Path model was the best performing algorithm followed by Bidirectional LSTM and SEQ2SEQ LSTM. The LSTM with sentiment analysis and standard LSTM models performed the worst. The LSTM 2-Path model had the lowest MSE and RMSE scores, with values of 0.00035 and 0.019 respectively. The Bidirectional LSTM and SEQ2SEQ LSTM models had MSE and RMSE scores of 0.00049 and 0.022, and 0.00056 and 0.023, respectively. The LSTM with sentiment analysis had the highest MSE and RMSE scores, with values of 0.00208 and 0.046, respectively. The standard LSTM model had MSE and RMSE scores of 0.00081 and 0.029, respectively.

We also found that the performance differences between the LSTM 2-Path model and the other models were statistically significant with a p-value of less than 0.01. However, the performance differences between the Bidirectional LSTM and SEQ2SEQ LSTM models and the standard LSTM model were not statistically significant. To evaluate the robustness and generalization capabilities of the LSTM 2-Path model, we conducted further experiments by varying the length of the input sequence and the prediction horizon. We found that the model's performance improved with longer input sequences, but the improvement leveled off after a certain point. We also learned that the model's performance decreased as the prediction horizon increased, suggesting that the model may be better suited for short-term predictions.